\documentclass[useAMS,usenatbib,usegraphicx]{mn2e}
\usepackage[english]{babel}
\usepackage{graphicx}
\usepackage{amsmath,amssymb,amsfonts}

\title[Primordial Anti-Biasing] {The Effect of Primordial Anti-Biasing on the Local Measurement of the 
Key Cosmological Parameters}
\author[Lee]
       {Jounghun Lee\thanks{E-mail: jounghun@astro.snu.ac.kr}\\
        Astronomy Program, Department of Physics and Astronomy, 
		   Seoul National University, Seoul 151-747, Korea}
\date{Accepted ***
      Received ***;
      in original form ***}
\pagerange{\pageref{firstpage}--\pageref{lastpage}} \pubyear{20??}

\begin{document}
\label{firstpage} \maketitle

\begin{abstract}
The best-fit values of the density parameter  and the amplitude of the linear density power spectrum 
obtained from the Cosmic Microwave Background (CMB) temperature field scanned by the Planck 
satellite are found to notably disagree with those estimated from the abundance of 
galaxy clusters observed in the local universe. Basically, the observed cluster counts are significantly 
lower  than the prediction of the standard flat $\Lambda$CDM model with the key cosmological 
parameters  set at the Planck best-fit values. We show  that this inconsistency between the local and 
the early universe can be well  resolved without failing the currently favored {\it flat} $\Lambda$CDM 
cosmology  if the local universe corresponds to a region embedded in a crest of the 
primordial gravitational potential field. Incorporating the condition of positive primordial 
potential into the theoretical prediction for the mass function of cluster  halos, we show that the 
observed lower number densities of the galaxy clusters are in fact fully consistent with the Planck 
universe.
\end{abstract}


\section{INTRODUCTION}\label{sec:intro}
The currently prevalent {\it flat} $\Lambda$CDM (Euclidean geometry, cosmological constant $\Lambda$  
and cold dark matter) cosmology can be completed only if the values of its key parameters are 
determined as precisely and accurately as possible.  Among the various probes that have so far been 
developed for the accurate measurements of the key cosmological parameters,  the best one is 
undoubtedly the Cosmic Microwave Background (CMB) temperature spectrum since the CMB sky 
provides us the least evolved version of our universe whose physics we believe are well understood.   
Although using the CMB spectrum as a cosmological probe suffers from the parameter degeneracy,  
this downside has been overcome by combining the CMB measurements with the results from the 
other complimentary probes such as the local cluster counts, the weak gravitational lensing, the 
baryonic acoustic oscillation (BAO) features and etc.

The recently reported  tensions between the parameter values determined from the local universe and 
from the CMB sky scanned by the Planck satellite \citep{planck_cp13} gave a great anxiety to the 
community, since the last (and perhaps the most crucial) puzzle to the standard flat $\Lambda$CDM 
picture is the consistency between the local and the distant scales (or the late and the early universe) 
\citep{tension13}. 
For instance, the Planck constraint on the Hubble constant, $H_{0}=67.3\pm 1.2$,  is lower than 
the locally determined value, $H_{0}=73.8\pm 2.4$,  by the Hubble Space Telescope (HST) 
observations of the Cepheid variable stars \citep{HST11}. 

The Planck constraints on the density parameter, $\Omega_{m}=0.314\pm 0.020$ , and the amplitude 
of the linear power spectrum, $\sigma_{8}=0.834\pm 0.027$, are also in significant tension with the local 
measurements, $\Omega_{m}=0.255\pm 0.043$ and $\sigma_{8}=0.805\pm 0.011$, from the low-$z$ 
cluster counts \citep{lowz_cl09}.  To make matters worse,  very recently, the Planck team 
traced the abundance evolution of the $189$ massive clusters detected via the Sunyaev-
Zel'dovich (SZ) effect and determined the best constraints as $\Omega_{m}=0.29\pm 0.02$ and 
$\sigma_{8}=0.77\pm 0.02$, which are even more serious $3\sigma$ deviations from the Planck 
best-fit ranges \citep{planck_sz13}.  These best-fit values of $\Omega_{m}$ and $\sigma_{8}$ determined 
from the late universe indicate that the observed number densities of the galaxy clusters in the local universe are 
much lower than predicted in the Planck cosmology. 

These tensions have caught immediate attentions, provoking a burst of research to find solutions as 
well as their origins.   Although some unknown systematics could have biased the local measurements 
of the cosmological parameters,  the recent hot trend is to suspect the model-dependent values of 
the Planck experiments and to suggest possible solutions based on such non-standard models as a 
$\Lambda$CDM with massive neutrinos \citep{nu_solve13}, a $\Lambda$CDM with scale dependent 
non-Gaussian initial conditions \citep{fnl_solve13},  a coupled dark energy (cDE) model  
\citep{cDE_solve13}, a $\Lambda$CDM with an "inhomogeneous geometry" \citep{swiss_cheese13} 
and so on. While the solutions to the tensions based on the above alternative models 
are definitely worth pursuing, it has to be noted that their capacity of alleviating the tensions have been 
achieved only at the cost of increasing the numbers of the cosmological parameters.  

In the current work,  we explore a possibility to explain away the tensions within the standard flat 
$\Lambda$CDM cosmology.  Our exploration will start from a core assumption that the local universe 
has formed in a crest of the primordial gravitational potential field and then will proceed in the direction 
of investigating if the observed lower amplitudes of the cluster mass 
functions are consistent with the Planck cosmology under this assumption.

\section{CLUSTER COUNTS IN A PRIMORDIAL POTENTIAL CREST}\label{sec:crest}

In the standard theory of structure formation, the primordial gravitational potential field, $\psi$, is 
regarded as a Gaussian random field, being related to the linear density contrast field $\delta$ as 
$\delta =\nabla^{2}\psi$.  \citet{LS98} demonstrated that since the primordial potential field is much 
smoother than the linear density field, it is possible to determine a  characteristic comoving scale, 
$R_{\psi}$,  of the "raw" unsmoothed primordial potential fluctuations as 
$R_{\psi}=(3\sigma^{2}_{\psi}/\sigma^{2}_{\nabla\psi})^{1/2}$  where the rms fluctuations of $\psi$ 
and ${{\bf\nabla}\psi}$ can be written in terms of the linear density power spectrum, $P_{\delta}(k)$, as 
\begin{equation}
\label{eqn:sigp}
\sigma^{2}_{\psi} =  \int_{k_{\rm nl}}^{\infty}\, dk\, k^{-2} P_{\delta}(k) \ , \qquad
\sigma^{2}_{{\bf\nabla}\psi} = \int_{0}^{\infty}\, dk\, P_{\delta}(k)\ ,
\end{equation}
where $k_{\rm nl}$ is the wave number corresponding the comoving cosmic horizon introduced to prevent 
a divergence of $\sigma^{2}_{\psi}$ in practical calculation. Note that there is no filtering by 
any kernel in Equation (\ref{eqn:sigp}), as stated in \citet{LS98}. Since the probability density of $\psi$ is 
Gaussian distributed,  finding a region with $\psi>0$ (crest) in the primordial potential field is as equally 
probable as that with $\psi<0$ (trough), i.e., $P(\psi>0)=P(\psi<0)=1/2$.  
   
Here, we set up a hypothesis that the present local universe have formed  in a primordial potential crest with $\psi>0$ 
rather than in a trough with $\psi<0$ in the early universe. This hypothesis naturally leads to an expectation that the
cluster number densities measured in the local universe would be lower than the global counterparts for a given 
background cosmology. The intriguing question is whether or not this hypothesis can explain away the tension between 
the local and the Planck measurements of the key cosmological parameters, especially, 
$\Omega_{m}$ and $\sigma_{8}$, on which the cluster number densities are most strongly dependent.

It is of importance to understand that the characteristic comoving scale $R_{\psi}$ of the primordial potential field 
represents its scale of coherence. 
According to \citet{LS98}, even though the formation of massive clusters are strongly biased toward the primordial 
potential troughs with $\psi<0$ \citep[see also][]{sahni-etal94,buriak-etal92,madsen-etal98,DD99}, it is not totally 
impossible for the massive clusters to form in a primordial potential crest.
Now, imagine a halo formed in a primordial potential crest with $\psi >0$. Since the peculiar velocity of this halo is 
always in the direction from a crest to a trough of the primordial potential field, it is expected that the halo would be 
displaced during the evolution by gravitational effect from the formation site. If its maximum displacement distance from 
the formation site is smaller than $R_{\psi}$, then this halo can be regarded as having effectively stayed in the primordial 
potential crest during the whole evolution. On the other hand, if its maximum displacement distance exceeds 
$R_{\psi}$, then the halo is regarded as having been displaced from the crest region in the subsequent evolution. 
As for the galaxy clusters, their maximum displacement distances are usually much smaller than  (see 
eq.[11] in Lee \& Shandarin 1998). Therefore, if formed in a primordial potential crest, the galaxy clusters 
are expected to have stayed in the crest during the whole evolution.
Unlike the smoothing scale, the characteristic scale $R_{\psi}$ is not an extrinsic scale but an intrinsic one 
determined only by the background cosmology. For the Planck cosmology with 
$\Omega_{m}=0.318,\ \Omega_{\Lambda}=0.683,\ n_{s}=0.962,\ \Omega_{b}=0.049,\ \sigma_{8}=0.834,\ h=0.671$ 
\citep{planck_cp13},  it is found to be $R_{\psi}\sim 100\,h^{-1}$Mpc. 

We adopt the analytic prescription laid out in the work of \citet{LS98} 
which basically incorporated  the effect of primordial potential on the number densities of the dark halos into 
the classical mass function formalism of  \citet[][hereafter, PS]{PS74}.  Although the PS mass function has 
been well known to be inaccurate when tested against the numerical results \citep[e.g.,][]{ST99,reed-etal03}, 
it is the only "purely" analytic mass function theory into which it is rather straightforward to incorporate the 
condition of $\psi>0$ without resorting to any empirical adjustment from N-body simulations. 
Since our goal here is not to model as accurately as possible the number densities of cluster halos in a 
primordial potential crest  but to see how much the condition of $\psi>0$ decreases the cluster number 
densities relative to the unconditional one, we believe that a modified version of the PS formalism should 
suffice to achieve this goal.

The PS formalism relates the differential mass function, $dN/dM$, of dark halos to the fraction of the 
volumes, $F(\delta_{c}; M)$, occupied by those regions whose linear density contrasts, $\delta$, exceed a 
unique threshold value $\delta_{c}=1.686$ \citep{GG72,peebles80,eke-etal96}, when the linear density field 
is smoothed on a given mass scale $M$:
\begin{eqnarray}
\label{eqn:dndm_ps}
\frac{dN}{dM} &=& 2\frac{\bar{\rho}}{M}\left\vert\frac{dF(\delta_{c},\ M)}{dM}\right\vert\ ,\\
\label{eqn:F}
F(\delta_{c}; M) &=& \int_{\delta_{c}}^{\infty}\, d\delta\, p(\delta; \sigma_{\delta})\ , 
\end{eqnarray}
where $p(\delta; \sigma_{\delta})$ is a Gaussian probability density of $\delta$ with the standard deviation 
$\sigma_{\delta}$, $\bar{\rho}$ is the mean mass density of the universe and the factor of $2$ before 
represents the normalization factor of the mass function \citep[e.g.,][]{bond-etal91,jedamzik95}. 
The standard deviation, $\sigma_{\delta}$, depends on the mass scale $M$ as
\begin{equation}
\label{eqn:sigd} 
\sigma^{2}_{\delta}(M) = \int_{0}^{\infty}\, dk\, k^{2} P_{\delta}(k) W^{2}(k;M)\ ,
\end{equation}
where $W(k;M)$ is the Fourier transform of a kernel by which the linear density contrast field is smoothed on 
the mass scale of $M$. 

Incorporating the condition of $\psi>0$ into the PS formalism amounts to modifying the volume fraction 
$F(\delta_{c};M)$ in Equation (\ref{eqn:F}) into 
\begin{equation}
\label{eqn:F_con}
F(\delta_{c};M |\psi>0) = \int_{\delta_{c}}^{\infty}\, 
d\delta\, p(\delta; \sigma_{\delta} ,\sigma_{c},\sigma_{\psi}\vert \psi >0)\ ,\
\end{equation}
with 
\begin{eqnarray}
\label{eqn:p_con1}
p(\delta, \sigma_{\delta},\sigma_{c},\sigma_{\psi}\vert\psi>0) &=& \frac{\int_{0}^{\infty}\, 
d\psi\,p(\delta,\psi;\sigma_{\delta},\sigma_{c},\sigma_{\psi})}{\int_{0}^{\infty}\,d\psi\,p(\psi;\sigma_{\psi})} \\
\label{eqn:p_con2}
&=& 2\,\int_{0}^{\infty}\, d\psi\,p(\delta,\psi;\sigma_{\delta},\sigma_{c},\sigma_{\psi})\ ,\
\end{eqnarray}
where $\sigma_{c}$ is the square root of the cross correlation between the {\it unsmoothed} primordial 
potential field, $\psi$, and the smoothed density field on the mass scale $M$, $\delta$. Since the Fourier 
transform of the unsmoothed primordial potential field $\tilde{\psi}$ is related to the Fourier transform of the smoothed 
linear density field $\tilde{\delta}$ as $\tilde{\psi}=k^{-2}\tilde{\delta}\,W^{-1}(k;M)$, the cross-correlation 
$\sigma^{2}_{c}$ can be evaluated as  
\begin{equation}
\label{eqn:sigc}
\sigma^{2}_{c}(M) = \langle \psi\,\delta\rangle = \int_{0}^{\infty}\, dk\, P_{\delta}(k) W(k;M)\ .
\end{equation}

\begin{figure}
\begin{center}
\centerline{\includegraphics[width=3.75in]{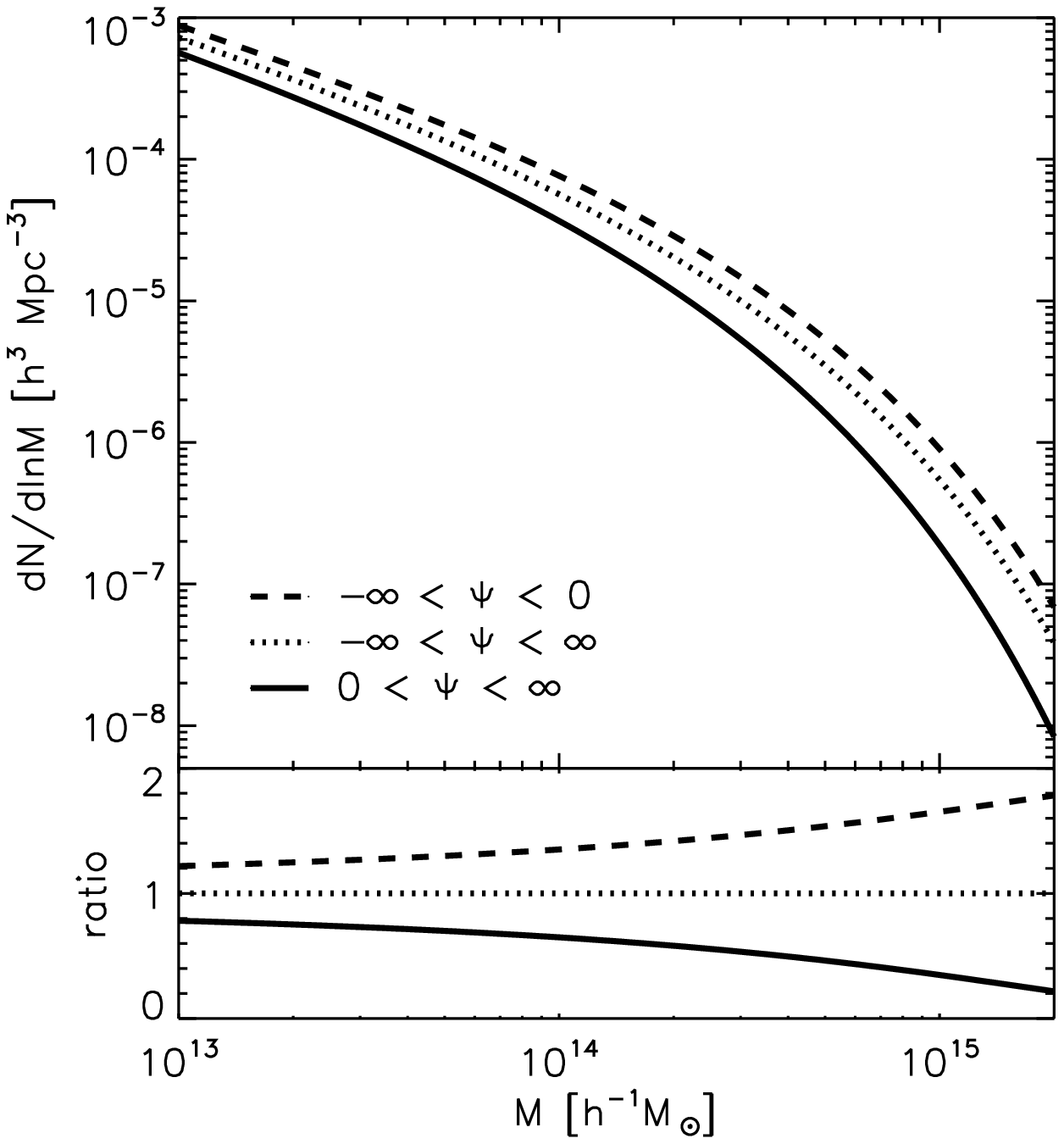}}
\caption{(Top panel): Conditional mass functions of the cluster halos at $z=0$ for the Planck 
cosmology ($\Omega_{m}=0.3175,\ \Omega_{\Lambda}=0.6825,\ h=0.6711,\ n_{s}=0.9624,\ 
\sigma_{8}=0.8344$). The solid  and dashed lines represent the anti-biased and the biased cases that 
the local universe corresponds to a crest and a trough of the primordial potential field, respectively. The 
case of no primordial bias, i.e., the unconditional mass function is plotted as dotted line for 
comparison. (Bottom panel): Ratio of the anti-biased (biased) case to that of no bias as solid (dashed) line. }
\label{fig:mf}
\end{center}
\end{figure}
The joint probability density, $p(\delta, \sigma_{\delta},\sigma_{c},\sigma_{\psi}\vert\psi>0)$, in Equation 
(\ref{eqn:p_con1}) can be straightforwardly calculated by using the statistics of a Gaussian random field 
\citep[][and references therein]{bbks86} as 
\begin{eqnarray}
\label{eqn:p_joint1}
p(\delta,\psi;\sigma_{\delta},\sigma_{c},\sigma_{\psi}) &=& 
\frac{1}{[(2\pi)^{2}\sigma^{2}_{\delta}(\sigma^{2}_{\psi}-\sigma^{4}_{c}/\sigma^{2}_{\delta})]^{1/2}}\times \\
&&\exp\left[-\frac{\delta^{2}}{2\sigma^{2}_{\delta}}-\frac{(\psi+\sigma^{2}_{c}\delta/\sigma^{2}_{\delta})^{2}}
{2(\sigma^{2}_{\psi}-\sigma^{4}_{c}/\sigma^{2}_{\delta})}\right]\, .
\end{eqnarray}
Now, let us perform a change of variable as
\begin{equation}
\label{eqn:nu_mu}
\nu \equiv \frac{\delta}{\sigma_{\delta}}\ , \qquad \mu\equiv  \frac{(\psi+\sigma^{2}_{c}\delta/\sigma^{2}_{\delta})^{2}}
{2(\sigma^{2}_{\psi}-\sigma^{4}_{c}/\sigma^{2}_{\delta})}\ .
\end{equation}
By applying the probability conservation relation of $p(\delta,\psi)d\delta\,d\psi = p(\nu,\mu)d\mu\,d\nu$ to 
$p(\delta,\psi)$, we derive the joint probability density distribution, $p(\mu,\nu)$, as
\begin{equation}
\label{eqn:p_numu}
p(\nu,\mu)d\nu\,d\mu =\frac{1}{(2\pi)^{1/2}}\exp\left(-\frac{\nu^{2}}{2}\right)
\frac{1}{(2\pi)^{1/2}}\exp\left(-\frac{\mu^{2}}{2}\right)\ ,
\end{equation}
As can be seen, the two variables, $\mu$ and $\nu$ are mutually uncorrelated and thus the joint distribution, 
$p(\mu,\nu)$,  is expressed as a product of two one-point distributions, $p(\mu)$ and $p(\nu)$.

Now, the volume fraction in Equation (\ref{eqn:F_con}) can be readily calculated in terms of $\mu$ and $\nu$ 
as
\begin{eqnarray}
\label{eqn:F_con2a}
F(M;\nu_{c},\mu_{c}) &=& 2\int_{\nu_{c}}^{\infty}d\nu\frac{1}{(2\pi)^{1/2}}\exp\left(-\frac{\nu^{2}}{2}\right)\times 
\\ \label{eqn:F_con2b}&&\int_{\mu_{c}}^{\infty}d\mu\frac{1}{(2\pi)^{1/2}}\exp\left(-\frac{\mu^{2}}{2}\right)\ ,
\end{eqnarray}
where $\nu_{c}=\nu(\delta=\delta_{c})$ and $\mu_{c}=\mu(\delta=\delta_{c},\psi=0)$.

Finally, the mass function of the cluster halos formed in a primordial potential crest is evaluated as
\begin{equation}
\label{eqn:dndm_con}
\frac{dN}{dM} = 2\frac{\bar{\rho}}{M}\left|\frac{dF}{dM}\right| 
=  2\frac{\bar{\rho}}{M}\left|\frac{d\nu_{c}}{dM}\frac{dF}{d\nu_{c}} 
+ \frac{d\mu_{c}}{dM}\frac{dF}{d\mu_{c}}\right| 
\end{equation}
where the four differentials can be computed with the help of the chain rule as
\begin{eqnarray}
\label{eqn:dfdnu}
\frac{dF}{d\nu_{c}}&=& \frac{1}{(2\pi)^{1/2}}\exp\left(-\frac{\nu^{2}}{2}\right)
\left[\frac{1}{2}{\rm erfc}\left(\frac{\nu_{c}}{2}\right)\right]\ , \\
\label{eqn:dfdmu}
\frac{dF}{d\mu_{c}}&=& \frac{1}{(2\pi)^{1/2}}\exp\left(-\frac{\mu^{2}}{2}\right)
\left[\frac{1}{2}{\rm erfc}\left(\frac{\mu_{c}}{2}\right)\right]\ , \\
\label{eqn:dnudm}
\frac{d\nu_{c}}{dM}&=& \frac{d\nu_{c}}{d\sigma_{\delta}}\frac{d\sigma_{\delta}}{dM}\ , \\
\label{eqn:dmudm}
\frac{d\mu_{c}}{dM} &=& \frac{d\mu_{c}}{d\sigma_{\delta}}\frac{d\sigma_{\delta}}{dM} + 
\frac{d\mu_{c}}{d\sigma_{c}}\frac{d\sigma_{c}}{dM}\, .
\end{eqnarray}
The top panel of Figure \ref{fig:mf} plots the {\it conditional} mass function, $dN/dM$, of the cluster halos per unit 
volume with $M\ge 10^{13}\,h^{-1}M_{\odot}$ at $z=0$ for two different cases. The solid and dashed lines correspond 
to the cases that the cluster halos formed in a primordial potential crest and trough, respectively.  The latter can be 
straightforwardly evaluated by repeating the same steps described in Equations 
(\ref{eqn:F_con})-(\ref{eqn:dmudm}) but with changing the crest condition of $\psi>0$ into the trough condition of 
$\psi<0$ as in \citet{LS98}. The dotted line in Figure \ref{fig:mf} corresponds to the case of no condition, i.e, the original 
{\it unconditional} PS mass function of the cluster halos.
The bottom panel shows the ratio of the two conditional mass functions to the unconditional one as solid 
and dashed lines, respectively, while the horizontal dotted line corresponds to unity. For this plot, the key 
cosmological parameters are set at the Planck values of $\Omega_{m}=0.3175,\ \Omega_{\Lambda}=0.6825,\ 
h=0.6711,\ n_{s}=0.9624,\ \sigma_{8}=0.8344$ \citep{planck_cp13}.  As can be seen, the number densities of the 
cluster halos formed in a primordial potential crest (trough) are indeed lower (higher) than the  unconditional 
counterpart. 
 
Since what has been always assumed in the theoretical modeling of the cluster mass function is that there is 
no difference between the local and the global average number densities of the clusters, our result implies that 
the comparison between the observed number densities of the local clusters and the analytic model of the unconditional 
mass function would yield different best-fit values of $\Omega_{m}$ and $\sigma_{8}$ from the Planck constraints, even 
when the background is truly the Planck universe.  
\begin{figure}
\begin{center}
\centerline{\includegraphics[width=3.75in]{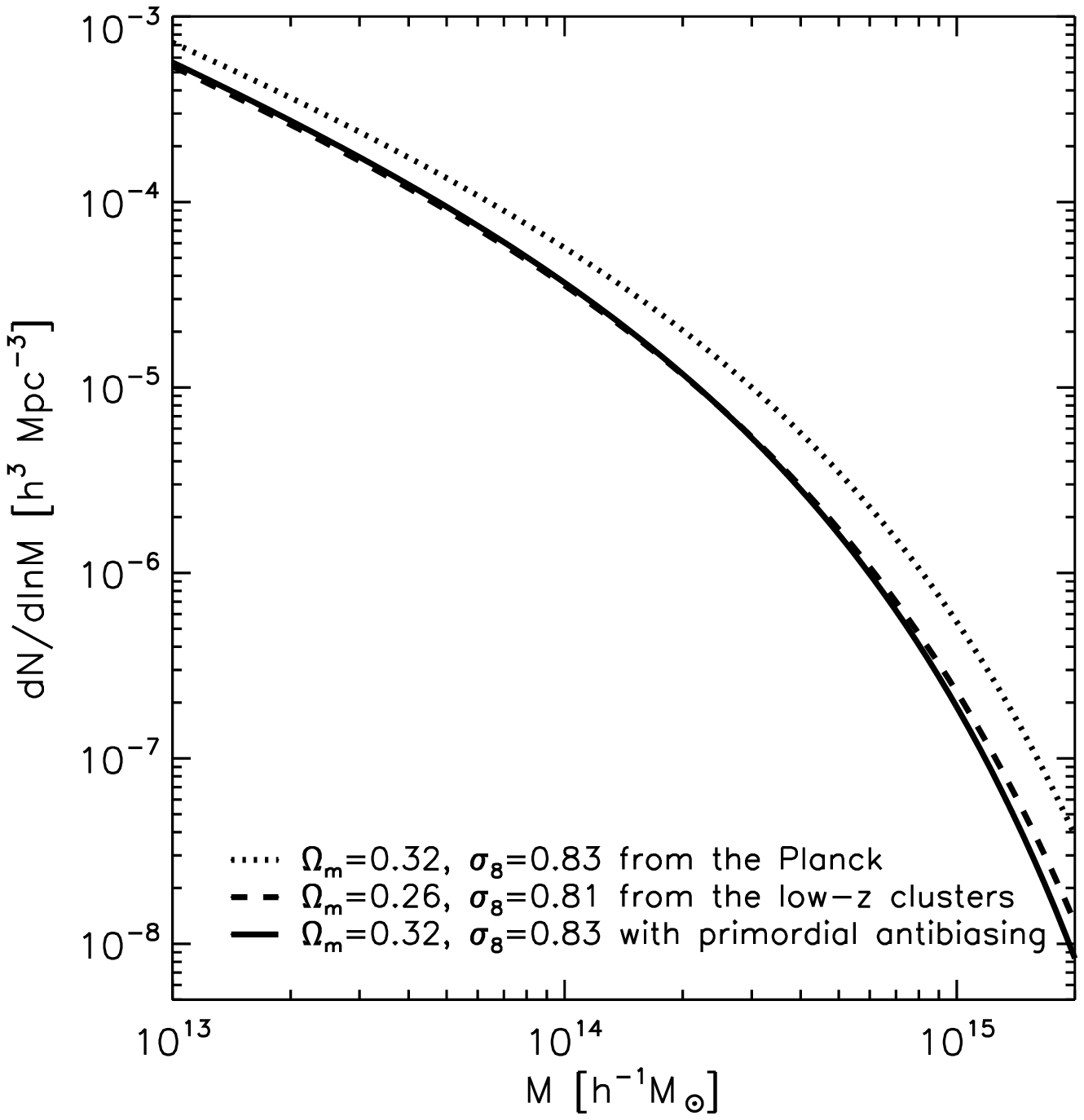}}
\caption{Unconditional mass functions of the cluster halos for two cosmologies: The dotted line 
corresponds to the case of the Planck cosmology with 
$\Omega_{m}=0.318,\ \Omega_{\Lambda}=0.683,\ h=0.671,\ \sigma_{8}=0.834$ \citep{planck_cp13}, 
while the dashed line corresponds to the low-$z$ cluster cosmology with 
$\Omega_{m}=0.255,\ \Omega_{\Lambda}=0.745,\ h=0.722,\ \sigma_{8}=0.805$ \citep{lowz_cl09}. 
The conditional mass function of the clusters formed in a primordial potential crest ($\psi>0$) for the 
Planck cosmology is plotted as solid line.}
\label{fig:mf_lowz}
\end{center}
\end{figure}

\begin{figure}
\begin{center}
\centerline{\includegraphics[width=3.75in]{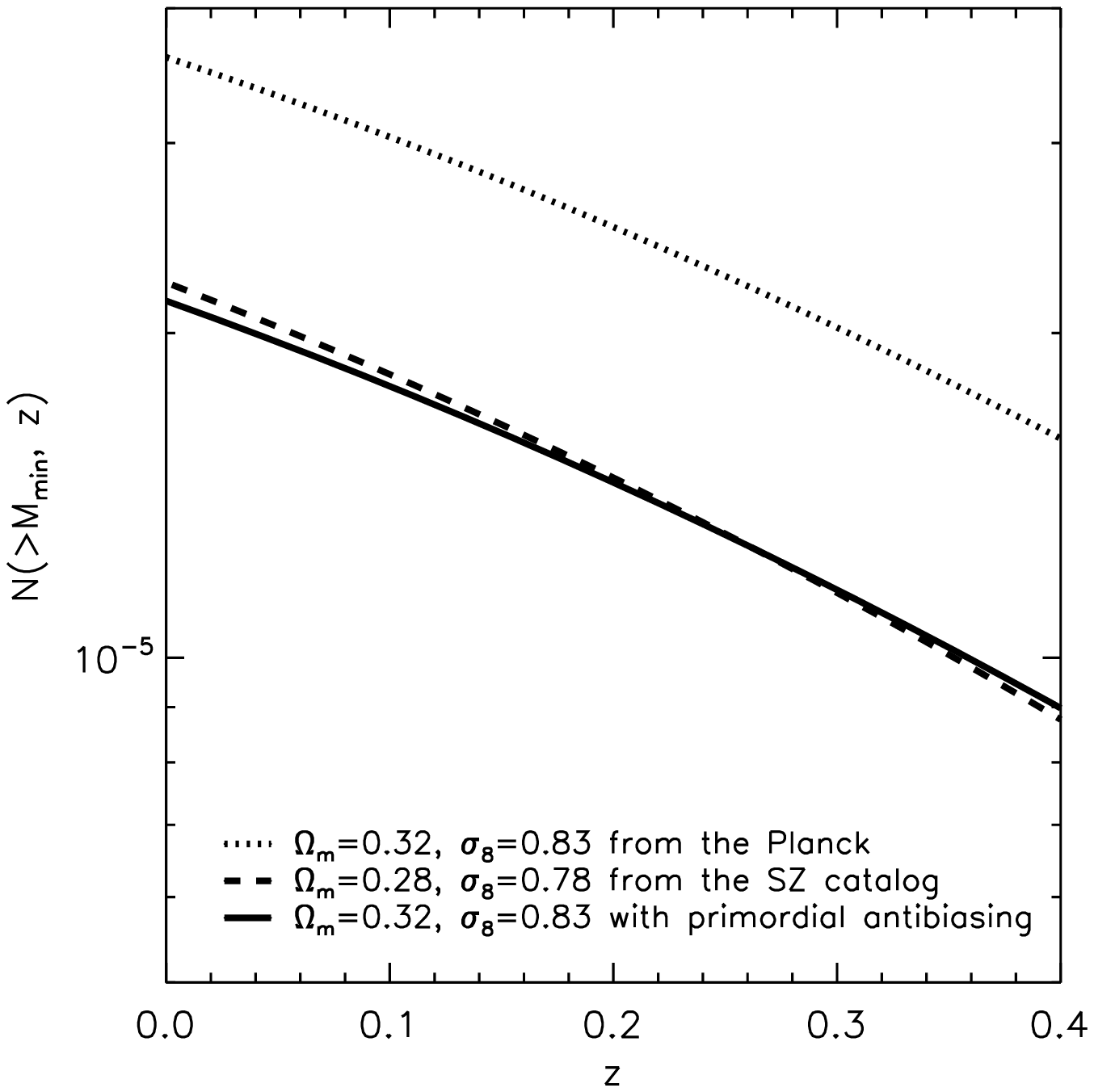}}
\caption{Unconditional number counts of the massive clusters with mass larger than 
$M_{\rm min}=10^{14}\,h^{-1}M_{\odot}$ as a function of redshift for two cosmologies: The dotted line 
corresponds to the case of the Planck cosmology with 
$\Omega_{m}=0.318,\ \Omega_{\Lambda}=0.683,\ h=0.671,\ \sigma_{8}=0.834$ 
\citep{planck_cp13}, while the dashed line corresponds to the cosmology from the Planck SZ catalog 
\citep{planck_sz13} with $\Omega_{m}=0.28,\ \Omega_{\Lambda}=0.72,\ h=0.722,\ \sigma_{8}=0.78$. The 
conditional number counts of the massive clusters formed in a primordial potential crest ($\psi>0$) for the 
Planck cosmology is plotted as solid line. }
\label{fig:mf_sz}
\end{center}
\end{figure}
We evaluate the unconditional  mass functions of the cluster halos by Equations (\ref{eqn:dndm_ps})-(\ref{eqn:F}) 
for two different cosmologies and plot them in Figure \ref{fig:mf_lowz} 
as dotted and dashed lines, respectively: The dotted line corresponds to the case of  the Planck cosmology 
with $\Omega_{m}=0.3175,\ \Omega_{\Lambda}=0.6825,\ h=0.6711, \sigma_{8}=0.8344$, while the dashed line 
corresponds to the case of the low-$z$ cluster cosmology with  
$\Omega_{m}=0.255,\ \Omega_{\Lambda}=0.745,\ h=0.722, \sigma_{8}=0.805$ \citep{lowz_cl09}. 
The comparison between the two cases obviously reveals that the unconditional cluster mass function for 
the low-$z$ cluster cosmology has a significantly lower amplitude than that for the Planck cosmology 
especially in the high-mass section. 

Evaluating the conditional mass function of the cluster halos formed in a primordial potential crest by 
Equations (\ref{eqn:F_con})-(\ref{eqn:dmudm}) for the Planck cosmology, we show it as solid line in 
Figure \ref{fig:mf_lowz}. Surprisingly, the solid line is in an excellent agreement with the dashed line. A crucial 
implication of this result is that the observed lower number densities of the local clusters are in fact fully consistent with 
the  Planck constraints as far as the condition of $\psi>0$ for the local universe is properly taken into account when the 
theoretical prediction for the cluster mass function is made. 

It is worth mentioning here that our analytic prescription of evaluating the conditional mass function of galaxy clusters in 
the local universe suffers from one ambiguity. Since we deal with the {\it scale-free} unsmoothed primordial potential 
field, the scale of its crest region in which our local universe is assumed to reside is unknown and thus has to be 
determined empirically.  Given that the effective redshift of the clusters considered by \citet{lowz_cl09} is 
approximately $0.15$, the scale of the local universe should extend at least up to the same redshift, $z=0.15$, 
which amounts to $430\,h^{-1}$Mpc for the Planck cosmology. In other words, to reconcile the Planck cosmology 
with the mass function of the low-$z$ clusters in the sample of \citet{lowz_cl09}, it has to be assumed that 
the local universe corresponds to a primordial potential crest of comoving size as large as 
$430\,h^{-1}$Mpc. 

Let us examine if incorporating the condition of $\psi>0$ can also resolve the other tension between the 
constraints of $\Omega_{m}$ and $\sigma_{8}$ from the Planck SZ catalogs  and the Planck cosmology. 
As mentioned in section \ref{sec:intro}, the abundance evolution of the SZ clusters measured by \citet{planck_sz13} 
has yielded the best-fit constraints of $\Omega_{m}=0.28,\ \sigma_{8}=0.78$ provided that the dimensionless 
Hubble parameter is given as $h=0.722$ (say, the SZ cluster cosmology). Here, the abundance evolution of the galaxy 
clusters, $dN/dz$, is proportional to the number counts, $N(>M_{\rm min},z)$, of the cluster halos as a function of 
redshift defined as $N(>M_{\rm min},z) = \int_{M_{\rm min}}^{\infty}\, dM\, dN/dM$, where $M_{\rm min}$ is the mass 
threshold set at $10^{14}\,h^{-1}M_{\odot}$ for our analysis. 

We evaluate  the {\it unconditional} number counts of the galaxy clusters as a function of redshift, 
$N(>M_{\rm min},z)$, by integrating Equation (\ref{eqn:dndm_ps}) over mass for the SZ cluster cosmology 
and show it as  dashed line in Figure \ref{fig:mf_sz}.  Comparing the dashed line 
with the dotted line in Figure \ref{fig:mf_sz} which is nothing but the unconditional number counts, 
$N(>M_{\rm min},z)$, for the Planck cosmology, one can see that the difference in the prediction for 
$N(>M_{\rm min},z)$ between the two cosmologies is quite large. 
Now, evaluating the conditional number counts, $N(>M_{\rm min}, z|\psi>0)$, by integrating 
Equation (\ref{eqn:dndm_con}) over $M$ for the Planck cosmology, we show it as solid line in Figure 
\ref{fig:mf_sz}.  It is exciting to see that the solid line exhibits a wonderful match (especially in amplitude) to the 
dashed line.

It is worth mentioning here that although the redshifts of the galaxy clusters in the Planck SZ catalog are distributed 
in a wide range of $0\le z\le1$, the constraints from the Planck SZ clusters on the values of $\sigma_{8}$ and 
$\Omega_{m}$ were decisively determined  in the the low-$z$ section ($0\le z< 0.4$). 
From Figure 7 in \citet{planck_sz13} which plots the number counts of the SZ clusters as a 
function of redshift, one can see that the data points in the high-$z$ section ($0.4\le z\le 1$) suffer from large 
uncertainties and that the amplitude of $dN/dz$ is determined almost decisively by those data points with 
small errors in the low-$z$ section ($0\le z<0.2$).  Thus, the anti-biasing effect of the primordial potential 
fluctuation can also reconcile the redshift distribution of the local SZ clusters with the Planck cosmology. 

\section{DISCUSSION AND CONCLUSION}\label{summary}
 
To solve the problem that the constraints on $\Omega_{m}$ and $\sigma_{8}$ from the observed cluster 
mass functions  in the local universe \citep{lowz_cl09} and from the redshift evolution of the SZ cluster 
counts \citep{planck_sz13} are both in tension with the constraints from the Planck experiments 
\citep{planck_cp13}, we have put forward a new hypothesis that the local universe formed in a crest of 
the primordial gravitational potential and has stayed in the potential crest after the formation. 
Under this hypothesis, we have evaluated analytically the number densities of the 
cluster halos formed in a primordial potential crest at present epoch with the help of the analytic 
prescriptions suggested by \citet{LS98} who had modified the original PS mass function theory to 
incorporate the effect of primordial potential.  

When the condition of being in a primordial potential crest is imposed,  the resulting {\it conditional} mass 
function of cluster halos has been found to exhibit significantly lower amplitude than the unconditional one.  
We have finally explained away the tension between the Planck and the local measurements of 
$\Omega_{m}$ and $\sigma_{8}$ by revealing the following two results: 
(i) The conditional mass function of cluster halos for the case of the Planck cosmology with 
$\Omega_{m}=0.318$ and $\sigma_{8}=0.834$ agree almost perfectly with the unconditional mass function 
for the low-$z$ cluster cosmology with $\Omega_{m}=0.255$ and $\sigma_{8}=0.805$; (ii) The  redshift 
evolution of the conditional  cluster counts for the case of the Planck cosmology matches that of the 
unconditional ones for the case of the cosmology from the Planck SZ clusters with $\Omega_{m}=0.28$ and 
$\sigma_{8}=0.78$.   

It should be worth discussing the apparent similarity and the essential difference between our analytic prescription of evaluating 
the cluster mass function  and the one proposed by \citet{alonso-etal12} in the context of the Lemairte-Tolman-Bondi cosmology   
\citep{lemaitre33,tolman34,bondi47}. Although both of the models have evaluated the halo mass function with an assumption that 
the local universe corresponds to a low-density region, the latter is based on the radical hypothesis that there is no dark energy in the 
universe \citep[see also][and references therein]{NS11} in a direct contrast to the former which is well accommodated by the 
standard $\Lambda$CDM cosmology.

Here we have followed the purely theoretical approach based on the analytic PS formalism of the halo mass function. 
However, it would be definitely desirable to make a direct comparison with the observed cluster counts, which requires 
to construct a more improved formalism for the conditional mass function, given the inaccuracy of the PS formalism 
\citep[e.g.,][]{ST99,reed-etal03}.  As for the unconditional mass function of cluster halos, there have been plenty of 
literatures which developed more improved analytic formalism by making more complicated and realistic 
assumptions about the halo formation process such as ellipsoidal collapse,  diffusive collapse threshold, and non-
Markovian random walks and etc \citep[e.g.,][]{SMT01,MR10,CA11,MS12,paranjape-eta12}.  
The formalism for the conditional mass function has to be improved and refined similarly before testing directly our 
hypothesis against observations. 

Although we have focused mainly on the parameters of $\Omega_{m}$ and $\sigma_{8}$ in the current work, 
it will be intriguing to investigate if our hypothesis can also lead to a reconciliation between the Planck and the 
local measurements of the Hubble constant $H_{0}$ \citep[e.g.,][]{HST11}. In fact, it can be logically expected that if 
the local universe corresponds to a primordial potential crest, then the density contrast averaged over the local 
universe would be lower than the global average, which would be reflected by a higher value of $H_{0}$ 
when measured locally than its global value determined from the CMB analysis.  
To quantitatively investigate this effect of primordial potential on the Hubble constant, however, it will be first 
required to study rigorously the evolution of the linear overdense region embedded in a primordial potential crest.

The other interesting issue that our hypothesis might be useful to address is the lower growth rate, $d\ln D/d\ln a$, 
inferred from the recently available deep galaxy surveys than predicted by the Planck cosmology
\citep[][and references therein]{lowerDz}.  Although the change of the linear growth factor, $D(z)$, by the primordial 
anti-biasing has not been accounted for in our analysis under the assumption that it would be small \citep{LS98}, the growth 
rate in a primordial potential crest that has a locally negative curvature must be lower than the average value. 
We believe that incorporating the condition of $\psi>0$ into the calculation of the linear growth rate might 
also provide a clue to resolving the apparent inconsistency found in \citet{lowerDz}.   
Our future work is in the direction of conducting these works.

\section*{Acknowledgments}

I thank the anonymous referee for helping me improve the original manuscript. 
This research was supported by Basic Science Research Program through the National Research Foundation 
of Korea(NRF) funded by the Ministry of Education (NO. 2013004372) and partially by the research grant from 
the National Research Foundation of Korea to the Center for Galaxy Evolution Research  (NO. 2010-0027910).


\end{document}